# Electron phase space control in photonic chip-based particle acceleration

R. Shiloh[1†], J. Illmer[1†], T. Chlouba[1†], P. Yousefi[1], N. Schönenberger[1, 2], U. Niedermayer[3], A. Mittelbach[1], P. Hommelhoff[1, 2]

[1]*Physics Department, Friedrich-Alexander-Universität Erlangen-Nürnberg (FAU), Staudtstraße 1, 91058 Erlangen, Germany*

[2]*Max Planck Institute for the Science of Light, Staudtstraße 2, 91058 Erlangen, Germany*

[3]*Technische Universität Darmstadt, Schlossgartenstraße 8, 64289 Darmstadt, Germany*

**Particle accelerators are essential tools in science, hospitals and industry[1–6]. Yet, their costs and large footprint, ranging in length from meters to several kilometres, limit their use. The recently demonstrated nanophotonics-based acceleration of charged particles can reduce the cost and size of these accelerators by orders of magnitude[7–9]. In this approach, a carefully designed nano-structure transfers energy from laser light to the particles in a phase-synchronous manner, thereby accelerating them[10]. However, so far, confinement of the particle beam in the structure over extended distance has been elusive; it requires complex control of the electron beam phase space and is mandatory to accelerate particles to the MeV range and beyond with minimal particle loss[11,12]. Here, we demonstrate complex electron phase space control at optical frequencies in the 225 nanometre narrow channel of a record-long photonic nanostructure. In particular, we experimentally show alternating phase focusing[11–14], a particle propagation scheme for, in principle, arbitrarily long minimal-loss transport. We expect this work to enable MeV electron beam generation on a photonic chip, with direct ramification for new forms of radiotherapy and compact light sources[9], and other forms of electron phase space control resulting in narrow energy or zeptosecond-bunched beams, for instance.**

Particle accelerators have enabled the discovery of the fundamental constituents of matter, exemplified by the recent identification of the Higgs particle[1,2]. In addition, they drive modern light sources such as synchrotrons and free electron lasers, which allow unrivalled insights to structural biology and material science[3,4]. Meter-long versions of accelerators provide radiation in virtually any modern hospital for cancer treatment[5,6].

It is common to accelerators that they accelerate charged particles while, at the same time, confining them transversally. This is challenging because the larger the acceleration, the larger the forces that disperse the beam, reflecting Liouville's theorem of phase space preservation[11,15,16]. In the currently employed radio-frequency (RF) accelerators methods have been found to accelerate and confine the particle beam[11].

Recently, laser-based electron acceleration has been demonstrated based on efficient momentum transfer from strong optical nearfields generated at nanophotonic dielectric structures[7,8] – first proposed in 1962[17,18] and today termed dielectric laser acceleration (DLA)[9]. It is the counterpart of classical RF acceleration, but at a factor of ~10,000 larger driving frequencies and, hence, in a structure with feature sizes smaller by the same amount. Intriguingly, at optical frequencies, dielectric structures can withstand peak fields roughly a factor of 100 larger than their metallic RF cavities counterparts. For this reason, DLA should facilitate accelerating gradients reaching 10 GeV/m.

Around 1 GeV/m has already been demonstrated[19]. Similarly, all functional elements required in any particle accelerator have now been demonstrated in DLA, including acceleration[7,8,19,20], deflection[21,22] and focusing[22–24] with purely optical forces, as well as waveguide-fed DLA[10].

The longest hitherto employed DLA structure for subrelativistic electrons was 13.2 µm long[25]. To greatly increase the acceleration length, active control schemes need to be devised to confine a beam and prevent particle loss[26]. We show here that with complex phase space control we can transport a beam through a 77.7 µm long structure. Without experiencing net acceleration, electrons are actively propagated through the DLA structure with the help of optical nearfield forces. Low-loss transport is of utmost importance because without it the best acceleration will not lead to noticeable electron currents, let alone to a power-efficient accelerator[9]. For transport, too, Liouville's theorem imposes severe limitations: Continuous focusing in all spatial directions would result in an increase of phase space density and is hence prohibited. Alternating phase focusing (APF) circumvents this by repeatedly alternating the focusing directions, and allowing defocusing in the other direction[11,13,14,18,27,28].

The force on the propagating electron can be written as[22]

$$F = \frac{qc}{\beta\gamma} \begin{bmatrix} \frac{1}{\gamma} C_C \sinh(k_x x) \cos\varphi_s \\ 0 \\ C_C \cosh(k_x x) \sin\varphi_s \end{bmatrix}, \quad (1)$$

where q is the electron charge, c the speed of light, β the velocity of the electron in units of c, γ the Lorentz factor, $C_C$ the excitation coefficient of the driving mode, $k_x$ the fundamental wave number of the DLA structure and $\varphi_s$ the synchronous phase (Fig. 1). We assume here that the particles propagate in the z direction, where x is the transverse direction parallel to the substrate (Fig. 2, 3), and that the particle velocity and the optical mode velocity are identical, i.e., the system operates in the phase-synchronous regime[11]. For this reason, no explicit time dependence shows up in Eq. 1, though it is implicitly reflected by the synchronous phase $\varphi_s$: The synchronous electron experiences a constant synchronous force, which depends strongly on the time the electron is injected into the optical mode (Fig. 1). The synchronous phase $\varphi_s$ and the incident laser phase have a fixed phase relation for purely periodic synchronous transport structures. However, the synchronous phase can be changed easily and almost instantaneously by introducing a gap in the periodic nanostructure, leading to a jump in the optical phase of the nearfield mode acting on the propagating electrons.

For an electron close to the central axis (x=0), we can plot the force components depending on $\varphi_s$ as depicted in Fig 1. For $\varphi_s = 3\pi/2$, the electron experiences a transversely focusing – longitudinally defocusing force (short: F phase), while for $\varphi_s = \pi/2$, the electron experiences a transversely defocusing – longitudinally focusing force (short: D phase). By variation of $\varphi_s$ we can repeatedly switch between the F and D phases (Fig. 2). The DLA structure is hence separated into F and D macro cells, realizing a FODO lattice, where O stands for no force (drift). It is well known that FODO lattices may lead to full confinement of the electron pulses[12]. Here, we choose the narrow transverse and the longitudinal direction as two π/2-out-of-phase directions. Because the nanophotonic pillars are 3 µm high (Fig. 3), we do not require forces along the y-direction (see supplementary information). Extension to a full three-dimensional confinement can be readily achieved if needed[27].

Fig. 2 shows the behaviour of electrons in transverse phase space (x-x') and longitudinal phase space (Δs – Δs') for one FODO lattice period. Here, x describes the particle position transversally off the centre line, and x' is the electron's angle with respect to straight propagation. Δs describes the deviation from the synchronous particle, given by a particle propagating at the synchronous phase, and

∆s' the longitudinal rate of change. Fig. 2 a-d show that when the electron is focused transversally it is debunched (defocused) longitudinally, and vice versa: Starting at z=0 ($\varphi_s = \pi/2$ in Fig. 2) the transverse size of the beam (x) decreases because x' is negative for positive x and vice versa (Fig. 2b) – the beam is converging. Simultaneously, the forces in this D macro cell counter this momentum, allowing the beam to reach a minimum at the centre of the macro cell before diverging again. After the phase jump to $\varphi_s = 3\pi/2$, the beam size (x) continues to grow due to x' being negative for negative x and vice versa. However, the optical force again counters this behaviour, allowing the beam to reach a point of maximum size in x before starting to focus again. This procedure repeats in each FODO lattice period. In parallel, the same happens in the longitudinal direction, but shifted in time (Fig. 2 c, d).

So far we have discussed the behaviour of an electron bunch occupying a small phase section to illustrate the dynamics. The experimental situation is more complex because the electrons arrive at the structure in an unbunched fashion[29], hence all phases are occupied. Yet, the APF effect in a pure transport structure works for all initial phases[14].

In addition, we have considered electron phase space control with an optical field strength perfectly matched to the particle beam and structure properties. For too small and too large optical forces, electrons will crash into the structure and are lost, which we will show in the following.

The photonic nanostructures used in the experiment consist of 3 µm high pillar pairs etched in mono-crystalline silicon with an open channel width of 225 nm (Fig. 3). The high-contrast structure comprises 10 macro cells, each consisting of 12 silicon pillar pairs. The required phase jump is realized with a 589 nm long drift gap between the macro cells. With this structure, we show successful beam transport over a total structure length of 77.7 µm, entailing 10 π-phase jumps, 10 macro cells, hence 5 FODO lattice periods.

The experimental setup is described in detail by Kozák et al[29]. Briefly, the electron beam is generated in a commercial scanning electron microscope (SEM) modified to produce a pulsed electron beam of ~600 fs duration at the interaction region, based on laser-triggering of the photocathode. The SEM allows us to focus the pulsed electron beam with an energy of 28.4 keV into the channel of the photonic nanostructure. After the interaction, the electron beam energy is measured with a magnetic spectrometer. The nanostructure is driven with femtosecond laser pulses of 1.93 µm central wavelength and 680 fs pulse duration, focused cylindrically to a spot size of 94 µm (1/e^2 radius) along the nanostructure channel and 14 µm along the pillar height. The laser field polarization points along the travel direction of the electrons, while the temporal overlap between laser and electron pulses is controlled by a delay stage in the laser beamline.

We measure the electron current through the high-contrast structure as a function of the peak optical field strength. With increasing laser intensity, the electron current increases steadily (Fig. 4a). At a peak field strength of 727.5 MV/m ± 73.6 MV/m, a factor of 2.67 ± 0.10 more electrons is transported through than without laser. Numerical particle tracking simulations, explained in detail in the supplementary information, fit the experimental behaviour perfectly well (solid line in Fig. 4a).

To show that a too large field strength leads to over-focusing and, hence, an again decreasing electron current, we would need to increase the field strength to above 800 MV/m in this structure, but for such large field strengths laser-induced damage sets in. Yet, we can design a structure with longer macro cell dimension so that optimal transport is achieved at a smaller field strength, at the cost of decreased total current transmission. This over-focusing structure is shown in Fig. 4 (f) and Fig. S3 in the supplementary information. Here we use 24 pillar pairs per macro cell such that the electrons experience the same phase (D or F) twice as long. This second type of APF structure contains 4 macro cells and two half cells at the beginning and the end, with a total length of 76.2 µm. The optimal transport field strength is 325 MV/m. For larger field strengths, the electrons are over-focused, resulting in a loss of particles, as observed in Fig. 4 (c). For peak fields larger than 550 MV/m, the normalized current falls below 1, indicating that now the laser fields actively destroy the beam rather than prevent it from crashing into the channel walls.

In Fig. 4 (b) we show an electron energy spectrum as a function of the temporal overlap between laser pulses and electron pulses in the high-contrast structure. For maximum overlap, the largest electron count rate is observed, which drops with decreasing temporal overlap. The spectral width remains minimal as predicted by our numerical simulations. Fig. 4 (d) shows the same temporal overlap scan for the over-focusing structure at 732 MV/m ± 70 MV/m peak electric field. Here we observe a minimum in the electron count rate. The advantage of the APF scheme is apparent: particles can be transported while confined in these narrow channels, all the while keeping the energy spread minimal.

Fig. 4 (e) depicts example simulated particle trajectories in the case of optimal guiding in the high-contrast structure. Some particles are lost right at in-coupling, but then the forces acting on the remaining particles are ideal to transport all of them through the structure. The simulated electron beam parameters are consistent with the experimental ones: The normalized input emittance was determined previously as 100 pm rad[29,30]. In contrast, Fig. 4 (f) shows an example of particle trajectories for the over-focusing case. Because the transverse forces acting on the particles are too strong, the focusing force is not matched to the phase jump intervals any longer, hence the electrons gain a too large transverse momentum and minimal-loss guiding cannot be maintained.

For both structure types, experimental results agree well with our particle tracking simulations. In addition to the transverse dynamics, the APF scheme inevitably also induces longitudinal phase space effects. They result from the selective energy modulation the electrons undergo as they are transported through the structure and manifest as bunching (Fig. 2 c, d). Because this is periodic with the pillar period of 619 nm, a train of sub-optical short bunches of electrons is generated every 3.23 fs (half an optical period). The generation of bunch trains of electrons can also be attained after propagation through a simpler uniform energy modulation structure plus ballistic propagation[25,31]. In that case, the electrons first gain a large energy spread and are then merely rotated by 90° in phase space. Particularly designed APF bunchers can compensate the created energy spread[14,32].

In conclusion, we have experimentally demonstrated ten-in-a-row well-controlled coupled transverse and longitudinal phase space rotations – at optical frequencies in a nanophotonic structure on a chip. We have implemented the alternating phase focusing scheme and could transport an electron beam over a length of 77.7 µm in a 225 nm wide channel, the aspect ratio of which corresponds to a length of ~7 m RF s-band particle accelerator. Because this scheme is scalable to, in principle, infinite length, we can now extend dielectric laser acceleration to long photonic structures. More importantly, we now have all ingredients demonstrated to create an APF accelerator structure, which could start with electrons at 30 keV and accelerate them up to 1 MeV. Assuming an average acceleration gradient of 1 GeV/m, 1 MeV is achieved on a 1 mm long chip. Such a device could result in compact light sources, potentially even compact free electron lasers, and sooner in miniature catheterized radiation therapy devices. Similarly, the high degree of phase space control will open the door to extreme electron beam parameters including narrow energy beams as well as bunching down into the zeptosecond realm[31,32].

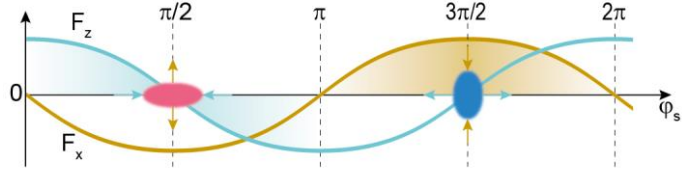

**Fig. 1. Forces acting as function of the synchronous phase $\varphi_s$.** The forces acting on a charged particle co-propagating with the optical nearfield mode according to Eq. 1, as a function of the particle's position inside of the mode, the synchronous phase $\varphi_s$. The longitudinal force $F_z$ is shown in cyan. The orange curve shows the transverse force $F_x$ acting on a particle at negative transverse (x) position. The force is flipped for positive x values, leading to focusing or defocusing behaviour, indicated by the arrows: Around $\varphi_s = \pi/2$, the pulse is transversally defocused but longitudinally focused (D phase), around $\varphi_s = 3\pi/2$, it is transversally focused but longitudinally defocused (F phase). The colour gradient indicates the strength of the respective focusing force.

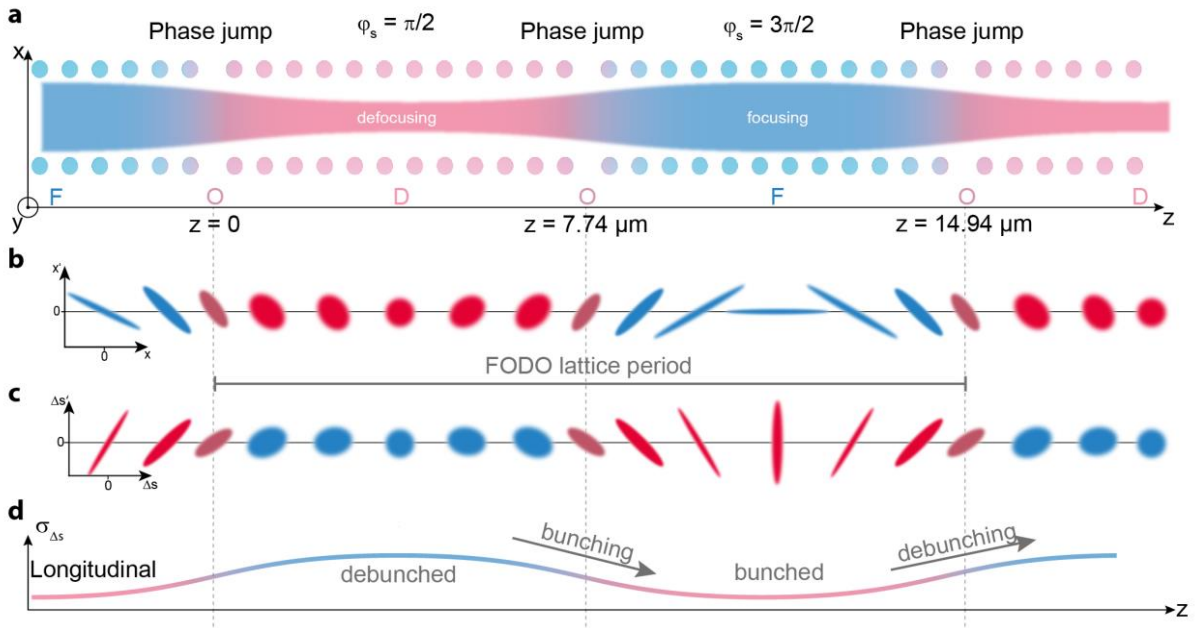

**Fig. 2. Complex optical electron phase space control in alternating phase focusing.** (**a**) Sketch of the electron beam envelope (coloured band) under the influence of the optical nearfield generated by the dielectric nanostructure. Filled circles indicate nanopillars extruded out of plane (y-direction). (**a, b**) Evolution of a particle ensemble in transverse phase space relative to the structure layout: The electron distribution at z=0 is compressed in transverse phase space (x extent decreasing), while in this D cell transversally defocusing forces already act on the electrons. The point of minimum transverse size is reached in the middle of the first macro cell (z = 3.87µm). At z = 7.74µm, a gap in the structure leads to a jump in the synchronous phase $\varphi_s$ between electrons and optical nearfield so that the particles now experience transverse focusing forces (F cell). (**c, d**) 90° out of phase relative to the transverse forces, the longitudinal forces act similarly on the longitudinal phase space distribution: the pulse length ($\sigma_{\Delta s}$) maximum in longitudinal direction coincides with the minimal extent in transverse direction and vice versa. The net effect is a confinement of the beam both transversally and longitudinally. For illustration purposes the particles depicted occupy a small initial phase volume.

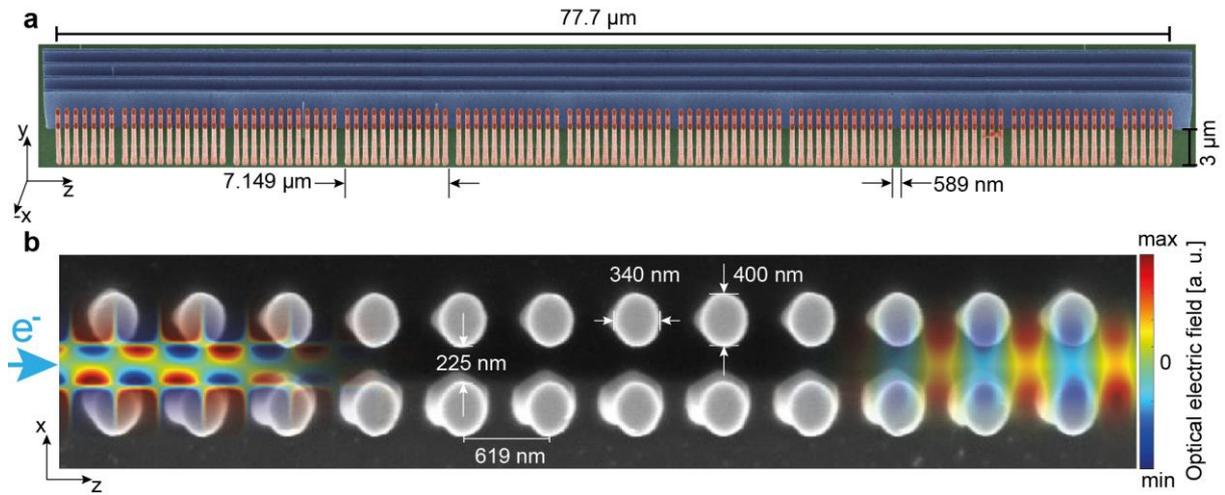

**Fig. 3. Silicon photonic phase space control nanostructure.** (**a**) Scanning electron microscope image of the 10 macro cell alternating-phase focusing structure (high-contrasts structure), coloured for clarity. Each full macro cell is 7.149 μm long and is separated from its neighbour cell by a 589 nm wide gap. The two half cells at the beginning and end ensure that an unbunched beam is optimally coupled in and out of the structure (see supplementary information). The total structure length equals 77.7 μm. The pulsed laser beam impinges on the nanostructure along the x-direction. A distributed Bragg reflector (blue) placed behind the pillar structure reflects the laser light back to create a symmetric field distribution[30]. (**b**) Top view of one macro cell. The pillars consist of elliptic silicon cylinders with the axes being 400 nm and 340 nm long, with a height of 3 μm. The distance between pillars is 619 nm. Dimensions have to be accurate to 20 nm to warrant efficacy of the structure (see supplementary information). Electrons (light blue arrow) are injected from the side. The colour superimposed on the left shows a snapshot of the electric field component exerting transverse forces (along x) on the electrons, while the colour on the right displays the electric field component in longitudinal z-direction.

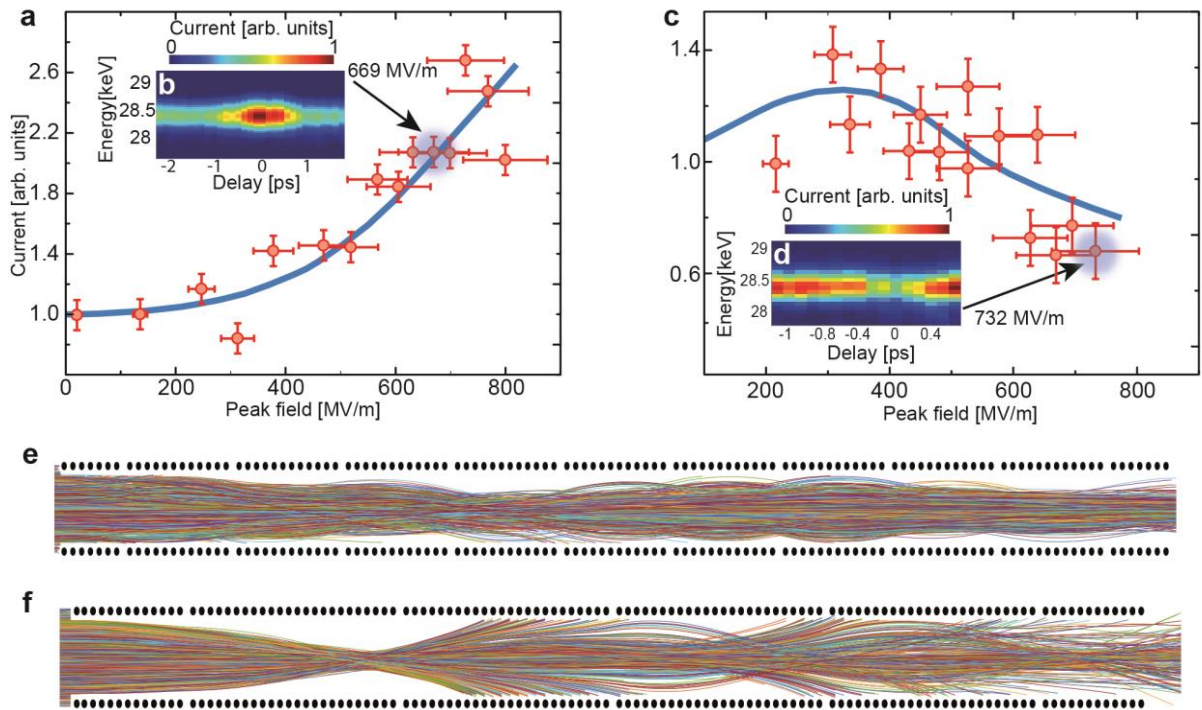

**Fig. 4. Experimental verification of the alternating-phase focusing scheme.** (**a**) Current after the high-contrast structure as function of the peak optical field. Red points: experimental data, blue curve: particle tracking simulation results. The current increases from 1 (laser off) with increasing field strength up to a maximum value of 2.67 ± 0.10. (**b**) Time delay scan between electron and laser pulses of the spectrally resolved current at 669 MV/m ± 64 MV/m. Clearly, the largest current is observed for maximum temporal overlap (zero time delay). (**c**) As in (a) but now for the over-focusing structure. Over-focusing sets in for fields larger than 300 MV/m. The current drops to below 1 for fields larger than 550 MV/m. (d) Same as in (b) for the over-focusing structure at 732 MV/m ± 70 MV/m. Now, the largest electron loss happens at maximum overlap. (**e**,**f**) Example particle trajectories in the high-contrast structure (e) and in the over-focusing structure (f). Clearly, excellent beam transport is observed in (e), while over-focusing is visible in (f). The colours in (e,f) are chosen to discern individual trajectories for presentation purposes.